\documentclass[english,10pt,aps,a4paper,preprintnumbers,floatfix,nofootinbib,showpacs,superscriptaddress,notitlepage]{revtex4-1}
\usepackage[english]{babel}
\usepackage[usenames,dvipsnames]{color}  
\definecolor{darkred}{rgb}{0.6,0,0}
\usepackage[colorlinks=true,citecolor=darkred,urlcolor=darkred, pdfborder={0 0 0}]{hyperref}
\usepackage{amsmath}
\usepackage{graphicx}
\usepackage{mathrsfs}   
\usepackage{amssymb}
\usepackage[normalem]{ulem} 
\usepackage{dcolumn}
\usepackage{bm}
\usepackage{multirow}
\usepackage{graphicx}
\usepackage[sort&compress]{natbib}
\usepackage{epstopdf}
\newcommand {\be} {\begin{equation}}
\newcommand {\ee} {\end{equation}}

\definecolor{greenLinks}{rgb}{0, 0.6, 0} 
\definecolor{blueLinks}{rgb}{0, 0, 0.6}
\definecolor{redLinks}{rgb}{0.6, 0, 0}
\definecolor{tempText}{rgb}{0.55, 0.10,0.67}
\definecolor{eprintLinks}{rgb}{0.4, 0.4, 0.4}
\definecolor{journalLinks}{rgb}{0.6, 0, 0}

\def\21{$\mathrm{SU(2)_L \otimes U(1)_Y}$ }
\def\31{$\mathrm{SU(3)_c \otimes U(1)_Q}$ }

\def\3211{$\mathrm{SU(3) \otimes SU(2)_L \otimes U(1)_R \otimes U(1)_{B-L}}$ }
\def\321{$\mathrm{SU(3) \otimes SU(2) \otimes U(1)}$ }
\def\422{$\mathrm{SU(4) \otimes SU(2) \otimes SU(2)_R}$ }
\newcommand{\LCDM}{$\Lambda$CDM}

\newcommand{\eV}{\mathrm{eV}}
\newcommand{\meV}{\mathrm{meV}}

\newcommand{\sumnu}{\Sigma m_\nu}
\newcommand{\ml}{m_\mathrm{lightest}}
\newcommand {\ignore}[1]{}

\newcommand{\sm}{{Standard Model }}

\newcommand{\AddrAHEP}{%
  AHEP Group, Institut de F\'{i}sica Corpuscular 
  (CSIC-Universitat de Val\`{e}ncia), Parc Cient\'ific de Paterna.\\
 C/ Catedr\'atico Jos\'e Beltr\'an, 2 E-46980 Paterna (Valencia) - SPAIN}

\begin{document}

\author{ Massimiliano Lattanzi}
\email{lattanzi@fe.infn.it}
\affiliation{Istituto Nazionale di Fisica Nucleare, sezione di Ferrara, Polo
  Scientifico e Tecnologico - Edificio C Via Saragat, 1, I-44122,
  Ferrara, Italy}
\author{Martina Gerbino}\email{gerbinom@fe.infn.it}
\affiliation{Istituto Nazionale di Fisica Nucleare, sezione di Ferrara, Polo
  Scientifico e Tecnologico - Edificio C Via Saragat, 1, I-44122,
  Ferrara, Italy}
\affiliation{HEP Division, Argonne National Laboratory, Lemont, IL 60439, USA}
\author{  Katherine Freese}\email{ktfreese@physics.austin.edu}
\affiliation{Physics Department, University of Texas, Austin, TX 78712}
\affiliation{The Oskar Klein Centre for Cosmoparticle Physics, Department of Physics, Stockholm University, SE-106 91 Stockholm, Sweden}
\affiliation{The Nordic Institute for Theoretical Physics (NORDITA), Roslagstullsbacken 23, SE-106 91 Stockholm, Sweden}
 \author{Gordon Kane} \email{gkane@umich.edu}
 \affiliation{Leinweber Center for Theoretical Physics, Department of Physics, University of Michigan, Ann Arbor, MI 48109, USA}
\author{Jos\'{e} W. F. Valle}\email{valle@ific.uv.es}
\affiliation{\AddrAHEP}

\title{\boldmath \color{BrickRed} Cornering (quasi) degenerate neutrinos with cosmology}


\begin{abstract}
\vspace{1cm}

In light of the improved sensitivities of cosmological observations, we examine the status of quasi-degenerate neutrino mass scenarios. 
Within the simplest extension of the standard cosmological model with massive neutrinos, we find that quasi-degenerate neutrinos are severely 
constrained by present cosmological data and neutrino oscillation experiments.
We find that Planck 2018 observations of cosmic microwave background (CMB) anisotropies disfavour quasi-degenerate neutrino masses at $2.4$ Gaussian $\sigma$'s, while adding Baryon acoustic oscillations (BAO) data brings the rejection to 5.9$\sigma$'s.
The highest statistical significance with which one would be able to rule out quasi-degeneracy would 
arise if the sum of neutrino masses
is  $\Sigma m_\nu = 60 \,\meV$ (the minimum allowed by neutrino oscillation experiments);
indeed a sensitivity of 15 meV, as expected from a combination of future cosmological probes, would further improve the rejection level up to 17$\sigma$.
We discuss the robustness of these projections with respect to assumptions on the underlying cosmological model, and also compare them with bounds from $\beta$ decay endpoint and neutrinoless double beta decay studies.

\end{abstract}

\maketitle

\medskip


\section{Introduction}
\label{sec:intro}


In analogy to the unification of the \sm gauge couplings~\cite{Georgi:1974yf}, it has been suggested that perhaps also the masses of the neutrinos may arise from a common seed at high energies. For example, a degenerate neutrino mass spectrum~\cite{Chankowski:2000fp} could emerge as a result of some SO(3) family symmetry~\cite{ioannisian:1994nx,Bamert:1994vc,
Ma:1998db,Barbieri:1999km,Antusch:2004xd} holding at high energies. Alternatively, a simple model with a discrete non-abelian symmetry A4 allows stacking the three lepton families as a triplet~\cite{babu:2002dz}, leading to quasi-degenerate neutrino masses. 
In this paper, we remain agnostic as to the underlying theory and consider forms of the mass matrix that could arise from a variety of models.

The goal of this work is to examine the consistency of such quasi-degenerate neutrino mass schemes in light of the improved sensitivity of cosmological observations, as well as improved and upcoming determinations of the neutrino oscillation parameters. 
We find that degenerate neutrinos are disfavoured by the combination of existing cosmological and oscillation data, being essentially ruled out in the case of inverted neutrino mass ordering, though still allowed within a relatively narrow region of parameters in the case of normal ordering. 

Several theoretical scenarios fit within the framework of the models we consider.
One simple scenario examined in this paper emerges as a result of the imposition of the A4 family symmetry to the \sm~\cite{babu:2002dz}.
It was originally proposed to provide a symmetry-based realization of the ``neutrino unification'' scenario suggested in~\cite{Chankowski:2000fp}. 
Within this picture the neutrino mass splittings needed in order to account for neutrino oscillation data~\cite{deSalas:2017kay,Esteban:2018azc} emerge as calculable radiative corrections~\cite{Hirsch:2003dr}. In its simplest form this scenario also predicts maximal atmospheric mixing, $\theta_{23} = \pi/4$, and vanishing reactor mixing angle, $\theta_{13} = 0$. However, reactor experiments~\cite{An:2012eh,Ahn:2012nd} have established that the reactor angle $\theta_{13}$ is non zero~(oscillation data such as those from T2K also indicate a growing hint in favor of leptonic CP violation). This implies that the simplest model needs amendment. Indeed, using the original picture as a ``zero-th order'' template, the scheme can be easily revamped in order to account for the required value of $\theta_{13}$. This makes the CP violating phase $\delta_{CP}$ physical, while at the same time providing stringent predictions in the $\delta_{CP}-\theta_{23}$ plane~\cite{Morisi:2013qna}. These will be testable within the upcoming generation of oscillation experiments~\cite{Chatterjee:2017ilf}. Our present dedicated cosmological analysis further constrains the parameters of the mass matrix for this specific scenario, making the degenerate neutrino scenarios strongly disfavoured.  

The paper is organized as follows. 
We first consider a general framework in which the small (solar) mass splitting is neglected, and write the neutrino mass matrix in a simple form that is representative of a wide class of theories.
This captures the most relevant features of degenerate schemes. We then extend our analysis to the more general scenario in which both atmospheric and solar mass splittings are taken into account. In all cases, we find that quasi-degenerate neutrinos are severely constrained by present cosmological data, at least in the simplest extension of the standard cosmological model with massive neutrinos. We also find that the case of inverted neutrino mass ordering is disfavoured. Future cosmological data would also rule out the surviving parameter regions still allowed for the quasi-degenerate normal ordered neutrino spectrum.  


\section{Preliminaries}
\label{sec:toy}

Before moving to the detailed description of the theory setup and the subsequent results, we would like to give some definition and then describe the bottom-line argument of our findings through some general considerations. Although these may fail in capturing the full complexity of the theory model to be discussed later, they serve to highlight the main line of reasoning that eventually will lead us to the results presented in this manuscript.

Strictly speaking, ``degenerate neutrino masses'' (or, for short, degenerate neutrinos) refers to the case in which the three masses $m_1,\,m_2,\,m_3$ are exactly equal, $m_1=m_2=m_3$, a possibility that is of course excluded by flavour oscillation experiments. In the following we will be concerned about constraining models of \emph{quasi-degenerate} neutrinos, meaning that the masses are only approximately equal, $m_1\simeq m_2\simeq m_3$. This approximate equality amounts to the requirement that the mass differences should be much smaller than the individual masses. In order to make quantitative arguments, like the one that follows, we need to set a criterion as to what ``much smaller'' means. For definiteness, we will define quasi-degenerate neutrino masses through the requirement that the largest mass difference should be a small fraction $\xi$ (with $\xi <1$) of the smallest individual mass.

The squared mass differences $\Delta m_{ij}^2\equiv m_i^2 - m_j^2$ 
are well measured in neutrino oscillation experiments. Global fits~\cite{deSalas:2017kay,Esteban:2018azc} to oscillation data yield  $\Delta m^2_{21}\simeq 7.6\times 10^{-5}\,\eV^2$ for the small (solar) squared mass splitting, and $|\Delta m^2_{31}|\simeq 2.5\times 10^{-3}\,\eV^2$ for the large (atmospheric) squared mass splitting. Since the sign of $|\Delta m^2_{13}|$ remains unknown, there are two possibilities for the ordering of neutrino masses: the so-called normal ($m_1 < m_2 < m_3)$ and inverted ($m_3 < m_1 < m_2$) orderings (NO and IO, respectively). The lowest value of the sum of neutrino masses $\Sigma m_\nu \equiv m_1 + m_2 + m_3$ allowed by oscillation measurements is $\Sigma m_\nu > 0.059\,\eV$ for normal hierarchy and $ > 0.10\,\eV$ for inverted hierarchy.  These minima are found by assuming the lightest neutrino is massless, and using the values reported above for the squared mass differences. However, since we are interested in the case of (quasi) degenerate neutrino masses, the neutrino masses must be substantially higher, so that the difference between the heaviest and lightest neutrino is smaller than any of the neutrino masses.

For the purposes of establishing a quantitative criterion for the definition of ``quasi-degenerate'', we may ignore the mass difference between $m_1$ and $m_2$, which is so much smaller than the mass difference between $m_1$ and $m_3$.  With this approximation $m_1 = m_2$, and a straightforward calculation shows that the criterion introduced above reads, for both NO and IO:
\begin{equation}
\frac{|m_3-m_1|}{\ml} \simeq \frac{\sqrt{\ml^2+|\Delta m^2_{31}|}-\ml}{\ml}\le\xi\,,
\label{eq:degcrit}
\end{equation}
where $\ml$ is the mass of the lightest neutrino, i.e. $m_1$ or $m_3$ for NO or IO, respectively.
Note that neglecting the solar mass splitting is appropriate for the purpose of verifying that the criterion for quasi-degeneracy is satisfied, since the value of the large mass difference $|m_3 - m_{1,2}|$ (where one should pick eigenstate $1$ or $2$ depending on the ordering) mainly depends on $|\Delta m^2_{31}|$.
Moreover,  in order to satisfy Eq.~(\ref{eq:degcrit}) with $\xi \ll 1$, the quantity $|\Delta m_{31}^2|/\ml^2$ should be much smaller than unity as well.
Expanding the square root in this limit,  Eq.~(\ref{eq:degcrit}) becomes
\begin{equation}
\frac{|\Delta m^2_{31}|}{2\ml^2}\le\xi\,,
\label{eq:degcrit2}
\end{equation}

Given the measured value of $|\Delta m^2_{31}|$, for a particular choice for the value of $\xi$, we can compute the smallest value of $\ml$ that satisfies the criterion as the value for which the equality in Eq.~(\ref{eq:degcrit}) or (\ref{eq:degcrit2}) holds. Taking $\xi = 0.1$ yields the condition $\ml > 0.11 \,\eV$. In terms of the sum of neutrino masses, a quantity well probed by cosmological observations, this corresponds to $\Sigma m_\nu \gtrsim 0.33\,\eV$.

As a result, an upper bound or a detection from cosmological arguments excluding $\Sigma m_\nu > 0.33\,\mathrm{eV}$ at high statistical significance, significantly reduces the parameter region where the neutrino mass spectrum can be degenerate. The latest bounds from the Planck satellite observations of the CMB anisotropies in temperature and polarisation, combined with measurements of the baryon acoustic oscillations (BAO), already corner the degenerate spectrum, providing $\sum m_\nu<0.12\,\mathrm{eV}$ at 95\% c.l.\footnote{Constraints obtained assuming a minimal one-parameter extension of the standard cosmological model, i.e. $\Lambda\mathrm{CDM}+\sum m_\nu$. A different cosmological model choice may result in a different bound on $\sum m_\nu$ from what is reported here. Further discussions on the choice of the cosmological model is given in Sec.~(\ref{sec:cosmology}).}. Further improvements are expected from the next generation of cosmological surveys, that will be able to reach a sensitivity of $\sigma(\sum m_\nu)\sim 15\,\mathrm{meV}$. In what follows, we will expand on this basic argument with a more articulated and thorough analysis.


\section{Theory setup}
\label{sec:setup}


In this section, we describe a theoretical setup that might be responsible for quasi-degenerate neutrino mass spectrum. To better illustrate the model, we start in Sec.~(\ref{subsec:setup1})
from a simple scenario in which the smallest (solar) neutrino mass splitting $\Delta m_{12}^2$ is neglected, reminiscent of the analysis of Ref.~\cite{Barbieri:1999km}.
We then move to Sec.~(\ref{subsec:setup2}), where we analyse the full-fledged scenario, with both the solar and atmospheric splittings taken into account.

\subsection{Simplest mass matrix}\label{subsec:setup1}

In this section, we begin by neglecting the smallest (solar) mass difference, i.e. we set $m_1=m_2$. We assume that the light neutrino mass matrix $\mathcal{M}_\nu$, possibly resulting from the seesaw mechanism, has the following form
\begin{equation}
\mathcal{M}_\nu = 
m_0 \left(
\begin{array}{ccc}
1+2\delta & 0 & 0\\
0 & \delta &1+\delta \\
0 & 1+\delta & \delta
\end{array}
\right)\, ,
\label{eq:Mnu1}
\end{equation}
where we use the weak eigenstate basis $(\nu_e,\,\nu_\mu,\nu_\tau)$, $m_0>0$ is the overall neutrino mass scale and $|\delta| \ll 1$ is a small real quantity responsible for the mass difference. This is one of the mixing patterns (specifically, the one called ``texture C'') appearing in Ref. ~\cite{Barbieri:1999km} (see Eq. 7 of that paper), once we identify
\begin{equation}
m_\mathrm{atm} \to m_0 \delta\, , \qquad M\to (1+2\delta)m_0\, ,
\end{equation}
where $m_\mathrm{atm}$ and $M\gg m_\mathrm{atm}$ introduced in Ref.~\cite{Barbieri:1999km} are the scale of atmospheric oscillations, and some higher mass scale, respectively. Notice that, to first order, $\delta = m_\mathrm{atm}/M$.

A pattern like the one in the mass matrix (\ref{eq:Mnu1}) might emerge, for example, in the presence of a non-Abelian family symmetry, either the small discrete A4 symmetry, or the continuous SO(3) symmetry. These could naturally lead to degenerate neutrinos, with small mass splittings induced as symmetry breaking effects. As an example, Ref.\cite{babu:2002dz} employs an $A4$ flavour symmetry, with calculable mass differences generated by radiative corrections (arising, for instance, in the context of softly broken supersymmetry). However, for the moment we will be agnostic and just assume that Eq.~(\ref{eq:Mnu1}) with $\delta \ll 1$ correctly describes the neutrino mixing pattern. We will, in any case, borrow some useful results from Ref. \cite{babu:2002dz}. 

Diagonalization of $\mathcal{M}_\nu$ yields the following exact positive eigenvalues:
\begin{subequations}
\begin{align}
&m_1 = m_0 \left|1+2\delta\right| \equiv m_0 \left| \eta \right| \, ,\\
&m_2 = m_0 \left|1+2\delta\right| \equiv m_0 \left| \eta \right| \, ,\\
&m_3 = m_0 \, ,
\end{align}
\label{eq:mass1}
\end{subequations}
where we have introduced the parameter $\eta \equiv 1+2\delta$, especially in view of the full-fledged scenario that will be discussed in the next section. Note that normal ordering ($m_3>m_1,\,m_2$) is realized for $|\eta|<1$, while inverted ordering ($m_3<m_1,\,m_2$) is realized for $|\eta|>1$. 
Since the mass eigenstates in Eqs.~(\ref{eq:mass1}) are independent of the sign of $\eta$, in our studies of cosmological bounds we can restrict just to the case $\eta \ge 0$. We can obtain a relation between $m_0$ and $\eta$:
\begin{equation}
m_0^2 = \frac{\Delta m_{31}^2}{1-\eta^2},
\end{equation}
where $\Delta m_{31}^2=m_3^2-m_1^2$ is well measured in oscillation experiments. The sum of the individual masses reads $\sum m_\nu = m_1+m_2+m_3 = m_0 (1+2\eta)$, a quantity well probed by cosmological observations.

From this one can derive a relation between $\sum m_\nu$ and $\eta$:
\begin{equation}\label{eq:2par}
\sum m_\nu = \left[\frac{\Delta m_{31}^2}{1-\eta^2}\right]^{1/2} (1+2\eta)
\end{equation}
Note that all results derived so far are exact and do not assume $\delta \ll 1$, i.e., $\eta$ close to unity. 

The quasi-degenerate case corresponds to $\delta \ll 1$ i.e. $\eta \sim 1$.
Given the measured value of $|\Delta m_{31}^2|$ from oscillations experiments, we have a relation between $\eta$ and $\sum m_\nu$, that allows one to translate cosmological constraints on $\sum m_\nu$ into constraints on $\eta$, and therefore on $\delta$.   In other words, using this relation we can use cosmological bounds on neutrino mass to strongly constrain the case of quasi-degenerate neutrinos.
The results of this analysis are reported below in Sec.~(\ref{subsec:results1}).

\subsection{The full mass matrix}
\label{subsec:setup2}

So far we have neglected the mass difference between the states $\nu_1$ and $\nu_2$ characterizing solar neutrino conversions. However, in the realistic case, one needs both mass splitting parameters $\Delta m_{21}^2$ and $\Delta m_{31}^2$ to be non-zero, in order to successfully describe the solar and atmospheric neutrino oscillation data. The generalization of the mass matrix in Eq.~(\ref{eq:Mnu1}) is given by \cite{babu:2002dz}:
\begin{equation}
\mathcal{M}_\nu = 
m_0 \left(
\begin{array}{ccc}
1+2\delta+2\delta' & \delta'' & \delta''\\
\delta'' & \delta &1+\delta \\
\delta'' & 1+\delta & \delta
\end{array}
\right)\, .
\label{eq:MnuRe}
\end{equation}
This mass matrix reduces to the one discussed in the previous section in the limit in which $\delta'$ and $\delta''$ are much smaller than $\delta$. 
For the moment we take all the parameters to be real. This simplifying approximation amounts to assuming that CP symmetry is conserved in neutrino oscillations, which is sufficient for our purposes~\footnote{Although there is evidence for CP violation in neutrino oscillations, e.g. from T2K data, cosmological observables are insensitive to it.}.
Moreover, without loss of generality we can again take $m_0>0$. 
The matrix ${\mathcal M_\nu}$ has positive eigenvalues given as
\begin{subequations}
\begin{eqnarray}
m_1&=&m_0 \left|1+2\delta+\delta'-\sqrt{\delta'^2+2\delta''^2}\right|,\\
m_2&=&m_0 \left|1+2\delta+\delta'+\sqrt{\delta'^2+2\delta''^2}\right|,\\
m_3&=&m_0. 
\end{eqnarray}
\label{eq:eigenvals}
\end{subequations}
One can see that quasi degenerate neutrino models correspond to the case where all three quantities are small: $\delta, \delta', \delta'' \ll 1$.\\
For convenience, let us define the following: 
\begin{eqnarray}
\eta&=&1+2\delta+\delta'\\
\eta'&=&\sqrt{\delta'^2+2\delta''^2}
\end{eqnarray}
so that the eigenvalues take the simpler form:
\begin{subequations}
\begin{eqnarray}
m_1&=&m_0 \left|\eta-\eta'\right|\, ,\\
m_2&=&m_0 \left|\eta+\eta'\right|\, ,\\
m_3&=&m_0  \, .
\end{eqnarray}
\label{eq:eigenvalseta}
\end{subequations}
Changing sign of either $\eta$ or $\eta'$ is equivalent to exchanging $m_1$ and $m_2$. Thus we restrict ourselves to the $(\eta > 0, \eta' >0)$ quadrant, where $m_1 < m_2$. Moreover, exchanging $\eta$ and $\eta'$ has no effect on the mass eigenvalues, so that in the rest of this section we can further restrict our attention to either of two octants; for definiteness, we take $\eta > \eta'$. 

Quasi-degeneracy corresponds to $\left|\eta-\eta'\right|\sim\left|\eta + \eta'\right|\sim 1$, which, in the first octant, 
 requires $\eta \sim 1$ and $\eta' \sim 0$.
 
Cosmological observations bound the sum of the individual masses, thereby placing stringent restrictions on these three parameters
\begin{equation}
\sum m_\nu=m_0\Big(\left|\eta-\eta'\right|+\left|\eta+\eta'\right|+1\Big) = m_0 \left(2 \eta+1\right),
\end{equation}
where the last equality holds when $\eta > \eta'$. 

As in the two-parameter approximation described in the previous section, one can use the results of neutrino oscillation experiments in order to express $\sum m_\nu$ as a function of $\eta$ only. 
To this aim, we first find the region of the $(\eta,\, \eta')$ plane that is consistent with current oscillation measurements. The requirement $m_2>m_1$, from the positive sign of the solar mass splitting, is always satisfied in the first quadrant. 
As explained above, in the following we consider the $\eta > \eta'$ region; the results can be easily extended to the rest of the parameter space from the symmetry arguments made above.  

\begin{figure}
\begin{center}
\includegraphics[width=0.48\textwidth]{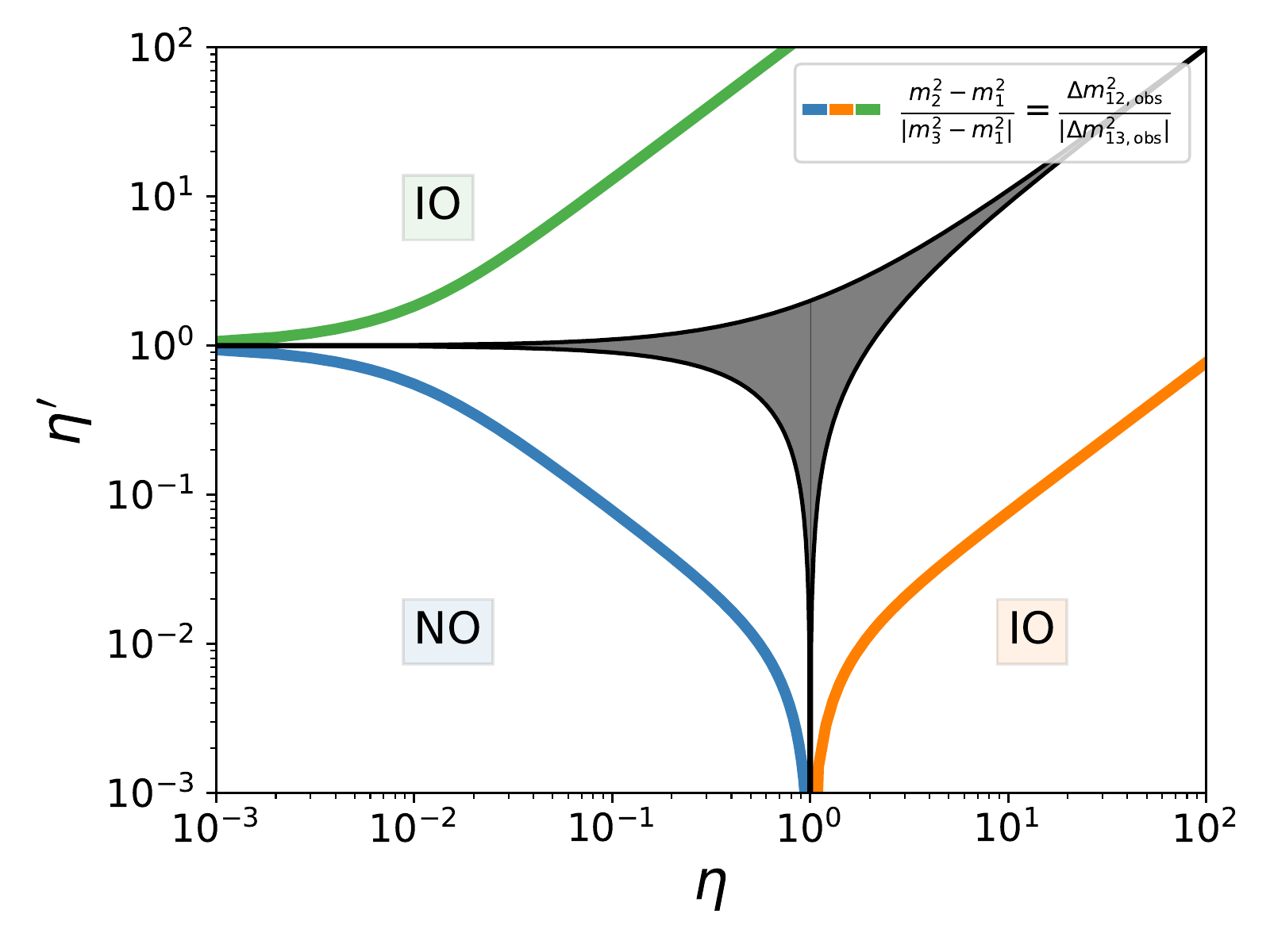}
\caption{Regions in the $\eta-\eta'$ plane, with $\eta > \eta'$, where normal or inverted ordering is realized, i.e. where the conditions $m_1<m_2<m_3$ (NO) or $m_3<m_1<m_2$ (IO) hold. There is one region for NO and two disconnected regions for IO. The central dark-shaded region is inconsistent with oscillation data as it corresponds to $m_1<m_3<m_2$. The solid curves are the loci of points that satisfy the constraint from neutrino oscillation measurements on the ratio $\Delta m_{12}^2/|\Delta m_{13}^2|$. The whole $(\eta,\,\eta')$ parameter space is symmetric for reflection around the bisector, i.e. $\eta \leftrightarrow \eta'$, so that one can focus only on the $\eta > \eta'$ region (see text for details).
}\label{fig:nhihosc}
\end{center}
\end{figure}

It is easy to verify that, when $\eta > \eta'$, normal ordering ($\Delta m_{13}^2,\,\Delta m_{23}^2>0$) is realized for $\eta' < 1-\eta$. Inverted ordering ($\Delta m_{13}^2,\, \Delta m_{23}^2 < 0$) is instead realized for $\eta' < \eta - 1$. The remaining region of parameter space should be excluded since there $m_1< m_3<m_2$ holds, inconsistent with oscillation experiments. These regions are shown in
Fig.~(\ref{fig:nhihosc}), where we have used symmetry arguments to reconstruct also the $\eta < \eta'$ part of the parameter space.

We can further impose that
\begin{equation}
\frac{m_2^2-m_1^2}{|m_3^2-m_1^2|}=\frac{\Delta m_{12,\mathrm{obs}}^2}{|\Delta m_{13,\mathrm{obs}}^2|}
\label{eq:oscconstr}
\end{equation}
where $\Delta m_{1x,\mathrm{obs}}^2$ for $x=2,3$ are the best fit values of the neutrino oscillation global fit analysis \cite{deSalas:2017kay}.  Solving 
Eq.~(\ref{eq:oscconstr}) with Eqs.~(\ref{eq:eigenvalseta}) provides a data-driven relation of the form $\eta'=\eta'(\eta)$. 
The relation is one-to-one separately in each of the two regions NO and IO in the first octant.
When the two regions are considered together, one finds that,
for a given value of $\eta'$, there are two values of $\eta$ that satisfy the oscillation constraints, one for each ordering. The curves that satisfy the constraint on $\Delta m_{12}^2/|\Delta m_{13}^2|$ are shown as solid thick lines in Fig.~(\ref{fig:nhihosc}), again after having been extended to the upper octant.
Note that the minimum value of $\eta$ in the first octant that satisfies the oscillation constraint~(\ref{eq:oscconstr}) is $\eta\simeq0.088$.

We have not yet fully exploited the information coming from neutrino oscillation experiments: given two measured mass differences, we have only required the theory to provide the observed ratio $\Delta m_{12}^2/|\Delta m_{13}^2|$. We can use the remaining information to express $m_0$ in terms of $|m_3^2-m_1^2| = |\Delta m_{31}^2|$. This finally allows us to express the three mass eigenvalues as functions of $\eta$ only. Thus we can write for the sum of the masses a generalization of Eq.~(\ref{eq:2par}):
\begin{equation}
\Sigma m_\nu = \left[\frac{\Delta m_{31}^2}{1-[\eta-\eta'(\eta)]^2}\right]^{1/2} (2\eta+1), \qquad (\eta > \eta') 
\label{eq:mnufull}
\end{equation}
where it is understood that we have used the constraints~(\ref{eq:oscconstr}) to express $\eta'$ as a function of $\eta$. The corresponding expression in the $\eta' > \eta$ part of the first quadrant can be obtained from Eq.~(\ref{eq:mnufull}) 
by the exchange $\eta \leftrightarrow \eta'$. Constraints on the mass parameters from cosmological data are presented in Sec.~(\ref{subsec:results2}).


\section{Cosmological data}
\label{sec:cosmology}


Here we describe in detail the cosmological data samples used to constrain the quasi-degenerate neutrino scenario. We employ the latest results published by the Planck collaboration on the sum of the neutrino masses $\sum m_\nu$~\cite{Aghanim:2018eyx}. These results were obtained assuming a $\Lambda\mathrm{CDM}+\sum m_\nu$ cosmological model. Such model was tested against the full suite of Planck satellite measurements of the CMB anisotropies in temperature and polarisation, and of the power spectrum of the gravitational lensing potential (a dataset combination labeled as ``TTTEEE+lowE+lensing'' in the Planck collaboration papers~\cite{Aghanim:2018eyx}),
combined with measurements of the BAO angular scale from 6dF~\cite{2011MNRAS.416.3017B}, SDSS-MGS~\cite{Ross:2014qpa} and SDSS-DR12~\cite{Alam:2016hwk} collaborations. We refer to this combination of data as ``Planck 2018 + BAO'' throughout the manuscript. In some cases, we will also present results for the Planck 2018 dataset alone. For the purposes of this work, we did not run the MCMC analysis. Instead, we have downloaded the MCMC chains provided by the Planck collaboration at the Planck Legacy Archive~\footnote{Chains available at this url:~\url{http://pla.esac.esa.int/pla}.} and reconstructed from them the posterior probability distribution of $\sum m_\nu$. Note that this means that we are implicitly assumimng a flat prior on $\sum m_\nu$. We make use of the posterior to derive the allowed regions for the parameters in the neutrino mass matrix, using standard statistical techniques. 

Future experiments will probe neutrino masses with higher sensitivity. We consider different combinations of future experiments as benchmarks for our projections. CMB observations from the Simons Observatory, combined with the large-scale CMB polarization data from Planck and measurements of the large-scale structure of the Universe, such as those from LSST~\cite{Mandelbaum:2018ouv}, DESI~\cite{DESI:2016tit}, Euclid~\cite{Amendola:2016saw} are expected to provide a sensitivity on $\sum m_\nu$, $\sigma(\sum m_\nu)\simeq30\,\mathrm{meV}$. This sensitivity can be improved to $\sigma(\sum m_\nu)\simeq20\,\mathrm{meV}$  by a cosmic-variance-limited measurement of the reionization optical depth $\tau$, such as that expected from the LiteBIRD satellite~\cite{Suzuki:2018cuy}. Finally, a sensitivity $\sigma(\sum m_\nu)\simeq15\,\mathrm{meV}$ is expected from the combination of ultimate CMB experiments, such as CMB-S4~\cite{Abazajian:2016yjj}, with LiteBIRD and the aforementioned  large-scale structure surveys.

To simulate the expected constraints from future cosmology, we interpret the sensitivity on $\sum m_\nu$ as the square-root of the variance of a Gaussian probability distribution centered in a given fiducial value of $\sum m_\nu$. We consider two different fiducial values of $\sum m_\nu$, which correspond to two detection scenarios: $\sum m_\nu=0.06\,\mathrm{eV}$ and $\sum m_\nu=0.1\,\mathrm{eV}$. These correspond to the case in which the ``true'' value of $\sum m_\nu$ is the minimal value allowed by measurements of the neutrino mass splittings from oscillation experiments in the normal and inverted ordering, respectively.  The highest statistical significance with which one would be able to rule out quasi-degeneracy would be for these cases of minimal allowed neutrino mass.
We make use of the Gaussian probability distributions so obtained to derive the expected allowed regions for the parameters in the neutrino mass matrix. The assumption of Gaussianity is expected to provide a good representation of the results from future cosmology, given the expected sensitivity of future surveys. 

A final remark concerns the choice of the underlying cosmological model. A well known limitation of the constraints from cosmological probes is the model-dependency issue, i.e. the fact that constraints on cosmological parameters may vary depending on the assumptions on the cosmological model. This happens because there is a certain level of correlation between different cosmological parameters. In other words, the physical effects of one parameter may be compensated by tuning other parameters. Such intrinsic uncertainty of the cosmological analysis can be cured in several ways providing confidence in cosmological results.  First of all, one can break the parameter correlation by combining observations of various cosmological probes (CMB and large-scale-structure) that depend differently on the same cosmological parameters. This is the reason why we adopt constraints from combined cosmological probes. Moreover, one can quantify statistically the preference for alternative cosmological models with respect to $\Lambda\mathrm{CDM}+\sum m_\nu$. To the best of our knowledge, there is no statistically significant preference reported for extended and/or exotic cosmological models that urges us to consider a different underlying parametrization than the one adopted in this manuscript~\cite{Gariazzo:2018meg}. Furthermore, the exquisite sensitivity and redundancy of future surveys will help further reduce the impact of model dependency~\cite{Brinckmann:2018owf}. See e.g. Ref.~\cite{Dvorkin:2019jgs} for a summary concerning the optimal combinations of future cosmological missions. This is the reason why, for the sake of simplicity, we choose to limit our analysis to the $\Lambda\mathrm{CDM}+\sum m_\nu$ scenario.However, in the conclusions we comment on the impact of considering different cosmological scenarios.


\section{Results of cosmological analysis}
\label{sec:results}

In this section, we report the main findings of our analysis. 
We use existing and upcoming bounds on the sum of neutrino masses from cosmology to examine the viability of the quasi-degenerate neutrino mass scenario.
We continue to study two cases: first, the ``simplest mass matrix" presented in Eq.~\ref{eq:Mnu1},
corresponding to  negligibly small solar mass splitting and, later, the more general mass matrix presented  in Eq.~\ref{eq:MnuRe}.

The basic approach is the following:  Taking advantage of relations we have previously found between $\sum m_\nu$ and parameters $\eta$ and $\eta'$ in
Eqs~(\ref{eq:2par}) and (\ref{eq:mnufull}) above, we will  convert cosmological bounds on $\sum m_\nu$, into bounds on $\eta$ and $\eta'$.  As a reminder, 
quasi-degenerate neutrino models (with small mass differences between species)
require $\eta \sim1$ for all the models we consider as well as $\eta' \sim 0$ for the more general mass matrix of Eq.~(\ref{eq:MnuRe}).
In the latter case, predictions for neutrino masses are unchanged if $\eta$ and $\eta'$ are exchanged, so that having $\eta\sim0$ and $\eta'\sim1$ also yields quasi-degenerate neutrino masses.
We will show that the combination of oscillations and cosmological bounds is essentially incompatible with such values of $\eta$ and $\eta'$.

In addition to studying how well quasi-degenerate neutrino mass can be ruled out from existing data, we also make projections for future data.
The highest statistical significance with which one would be able to rule out quasi-degeneracy would be for the case of minimal neutrino mass allowed by oscillations data, i.e.
$\Sigma m_\nu = 0.06 (0.1) $ eV for NO (IO). 
Specifically, we examine the bounds on quasi-degeneracy with sensitivity of upcoming experiments  ranging from $\sigma\left(\Sigma m_\nu \right) = 0.030\,\eV$ to  $\sigma\left(\Sigma m_\nu \right) = 0.015\,\eV$, assuming the 
 sum of neutrino masses is this minimal allowed value\footnote{If the actual sum of neutrino masses is higher than these minimal values, while still being consistent with bounds from cosmology, then the quasi-degenerate scenario would still be ruled out albeit at a slightly lower statistical significance.}.

\subsection{Results in the simplest mass matrix model}\label{subsec:results1}

Current cosmological data from Planck place limits on the parameters describing the quasi-degenerate neutrino scenario. Within the approximation where the solar neutrino mass splitting is
neglected (see Sec.~(\ref{subsec:setup1})), the neutrino mass matrix is given in Eq.~(\ref{eq:Mnu1}), with
the parameters $\delta$ and $\eta$ as defined in Eq.~(\ref{eq:mass1}).  In this simplest case, we have only two parameters, $\sum m_\nu$ and $\eta$, related 
by Eq.~(\ref{eq:2par}); the relation between them depends on $\Delta m_{31}^2$, a quantity measured by oscillation experiments, 
%
 
\begin{figure}
\begin{center}
\includegraphics[width=.7\textwidth]{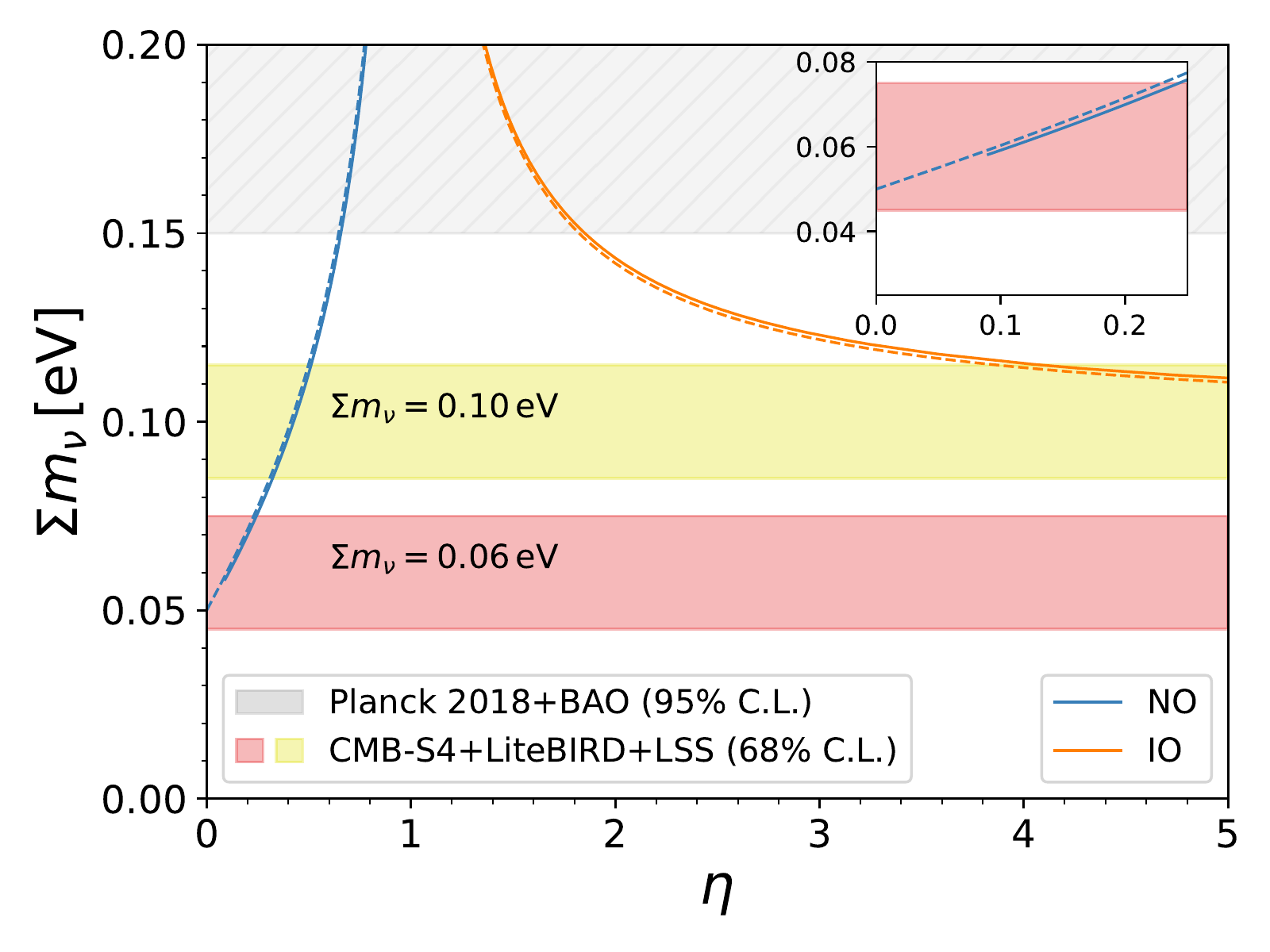}
\caption{Bounds from existing and upcoming cosmological observations on the sum of the neutrino masses $\sum m_\nu$ as a function of the parameter $\eta$ for the ``simplest mass matrix''
(Eq.~(\ref{eq:2par}), dashed curves) and ``full mass matrix'' (Eq.~(\ref{eq:mnufull}), solid curves), given the constraints from oscillation experiments, for NO (blue curves) and IO (orange and green curves).
The color code matches the one used in Fig.~\ref{fig:nhihosc}. For the simplest mass matrix, there are two curves, one for each ordering. In the full mass matrix case, we assumed $\eta>\eta'$,
  always possible without loss of generality, see discussion in the text. With this choice, we have two branches for each ordering also in the full mass matrix case. Note that oscillation experiments require that $\eta \ge 0.088$, corresponding to $\sum m_\nu = 0.06\,\eV$, the minimum mass still allowed for NO given the best-fit of current oscillation data. This is shown in the small inset on the top right. The minimal mass IO scenario $\sum m_\nu = 0.10\,\eV$ is instead realized for $\eta\to \infty$. Quasi-degenerate masses are obtained for $\eta \simeq 1$, which requires $\sumnu$ to be large ($\sumnu \to \infty$ as $\eta\to 1$).
  The grey hatched region $\sum m_\nu > 0.15\,\eV$ is excluded by present cosmological (Planck 2018 + BAO) and oscillation data at 95\% level. The two horizontal bands show the 68\% credible interval for $\sum m_\nu$ from future experiments given $\sum m_\nu = 0.06\,\eV$ or $\sum m_\nu = 0.10\,\eV$, corresponding to the minimal possible masses for NO and IO, and a sensitivity $\sigma\left(\Sigma m_\nu \right) = 0.015\,\eV$ (see text). The grey region and the colored bands serve as rough guides to the constraining power of present and future data.   
  See text and tables for information about credible intervals for $\eta$.}
\label{fig:sumnu_vs_delta_noapprox}
\end{center}
\end{figure}

 Fig.~\ref{fig:sumnu_vs_delta_noapprox} shows our results for the simplest mass matrix model in the $\sum m_\nu - \eta$ plane.  Here the dotted curves show the sum of the neutrino masses as a function of the parameter $\eta$ through Eq.~(\ref{eq:2par}), where the value of $\Delta m_{13}^2$ is fixed at the best-fit value from global fits of neutrino oscillation experiments.  The dotted blue curve corresponds to normal ordering while the dotted orange curve corresponds
to the inverted ordering. 
We recall that  in the simplest mass matrix model, $\eta>1$ yields inverted ordering, while $\eta<1$ gives normal ordering.

Cosmological bounds are indicated in Fig.~\ref{fig:sumnu_vs_delta_noapprox} by horizontal bands of various colors. 
The gray-hatched region is excluded by the current cosmological data (Planck 2018 +BAO). Note that the upper bound on $\sum m_\nu<0.12\,\mathrm{eV}$ quoted in the Planck 2018 parameters paper has been derived assuming a prior $\sum m_\nu >0$. 
In the present analysis, we are using information from oscillation experiments that require $\sum m_\nu >0.06 \,\eV$, so we should use a prior that reflects this knowledge. Taking this into account yields $\sum m_\nu<0.15\,\mathrm{eV}$ at 95\% c.l., which is the value used to produce the gray-hatched exclusion region in Fig.~\ref{fig:sumnu_vs_delta_noapprox}.
Since quasi-degenerate neutrino masses require $\delta\ll1$ ($\eta \sim 1$), one can see already by eye that this scenario is ruled out. Roughly, one can see that 
 $\eta \gtrsim 1.8$ is required to satisfy the cosmological bounds for inverted ordering, and $\eta\lesssim0.7$ for normal ordering.
  This range will be further reduced once data from future cosmological surveys become available. The expected sensitivities of future cosmology are shown as colored horizontal bands for the two fiducial values of $\sum m_\nu$ introduced in Sec.~(\ref{sec:cosmology}). These have been chosen as the minimal masses allowed by neutrino oscillations, and correspond, for each ordering, to the strongest rejection for quasi-degeneracy.
The viable region of the quasi-degenerate model reduces to the ranges in which the lines overlap with the colored bands.

\subsection{Cosmological bounds on neutrino mass in the full mass matrix approach}
\label{subsec:results2}

We now move to the full theory setup described in Sec.~\ref{subsec:setup2}.
There, we have shown how we can use neutrino oscillation data to express the three mass eigenvalues, and their sum, as functions of $\eta$ only, see Eqs. (\ref{eq:eigenvalseta}) and (\ref{eq:mnufull}). 
In the rest of this section we will further assume, unless otherwise stated, that $\eta \ge\eta'$, which implies $\eta \ge 0.088$ once oscillation data are taken into account. From the discussion in Sec.~\ref{subsec:setup2}, it is clear that one can make this choice without loss of generality.
Similarly to what we have done in Sec.~(\ref{subsec:results1}) for the simplest model, we show in Fig.~(\ref{fig:sumnu_vs_delta_noapprox}) the sum of neutrino masses $\sum m_\nu$ as a function of $\eta$.
The $\sum m_\nu(\eta)$ relation can be used to translate cosmological constraints on $\sum m_\nu$ into constraints on $\eta$. This operation requires some care given the two-valued nature of the $\sum m_\nu-\eta$ relation and the multimodality of the posterior. In this section we only report our results; the interested reader is referred to the appendix for technical details on how probabilities and exclusion levels are computed. We only report results in terms of $\eta$. Credible intervals for $\eta'$ can be obtained using the $\eta'(\eta)$ relation built as explained in Sec.~\ref{subsec:setup2}.

\begin{figure}
\begin{center}
\includegraphics[width=.45\textwidth]{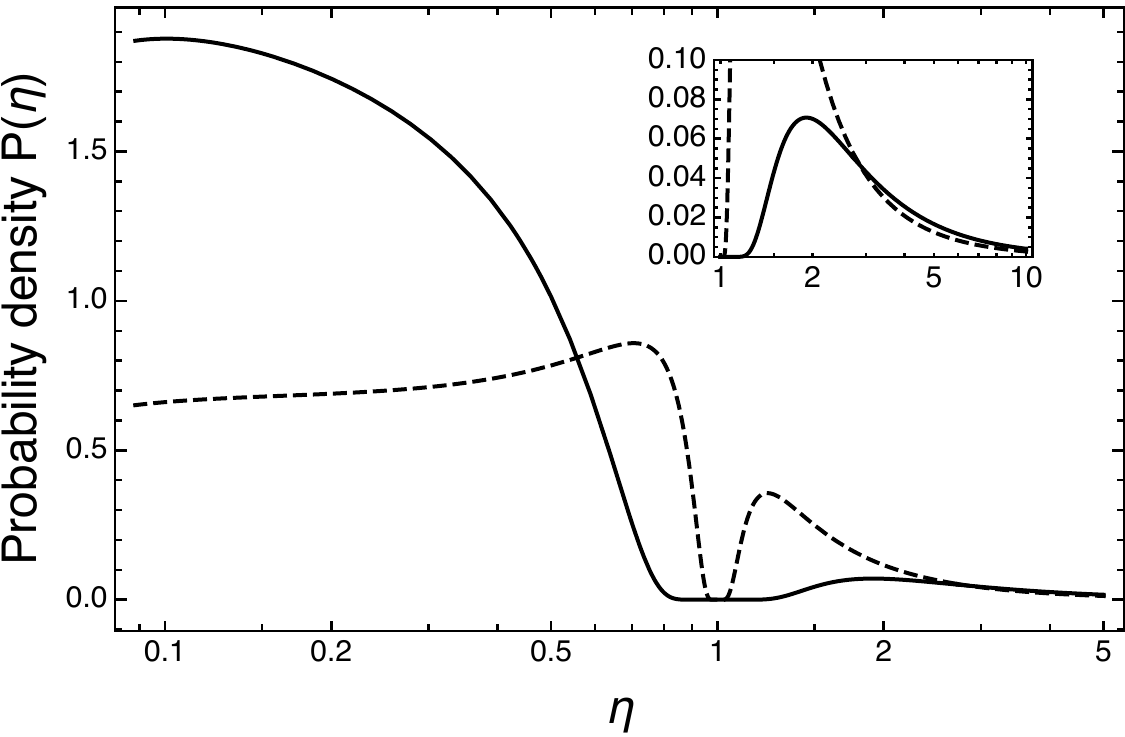}
\caption{Posterior probability density for $\eta$ given current cosmological and oscillation data. Note the logarithmic scale on the horizontal axis. The solid curve uses the Planck 2018+BAO as the cosmological dataset, while the dashed curve is for Planck 2018 only. The inset shows a blow-up of the region $\eta >1$. Note also the different scale in the vertical axes between the main panel and the inset. The region $\eta<1$ ($\eta>1$) corresponds to normal (inverted) ordering. Quasi-degenerate neutrino masses are realized for $\eta \simeq 1$.}
\label{fig:post_eta}
\end{center}
\end{figure}

We show the posterior for $\eta$ from Planck2018 and Planck 2018+BAO in Fig.~(\ref{fig:post_eta}), highlighting the multimodality of the probability distribution.
The 95\% credible intervals on the parameter $\eta$ (with $\eta \ge \eta'$) are reported in Tab.~\ref{tab:etafut2}, for different combinations of current and future cosmological datasets. It is clear that the quasi-degenerate case, corresponding to $\eta\simeq 1$ is strongly disfavoured by the data. It is useful to quantify the preference of the data for non-degenerate neutrinos. To this purpose, we define the quasi-degenerate scenario as the one in which the large mass splitting is smaller than 10\% of the overall mass scale.
From Planck 2018+BAO data, we get $P_\mathrm{deg} = 4\times 10^{-9}$, corresponding to $5.9$ Gaussian $\sigma$'s in favour of nondegenerate neutrinos. See the appendix for more details on how $P_\mathrm{deg}$ is defined and computed. The preference is relaxed to $2.4 \,\sigma$'s (odds of $64:1$) if only Planck data are considered. In the last column of Tab.~\ref{tab:etafut2}, we also report values of the Bayes factor $B$ between the quasi-degenerate and non-degenerate scenarios, defined as $B \equiv P_\mathrm{deg}/(1-P_\mathrm{deg})$ (see appendix for more details).

Since any given pair $(\eta, \eta')$ allowed by oscillation experiments uniquely corresponds to either NO or IO (see Fig.~\ref{fig:nhihosc}, the posterior for $\eta$ can be used to assess preference for one ordering or the other. We find that Planck 2018 + BAO prefers normal over inverted ordering with odds $3.3 : 1$ ($1.2 \sigma$'s).

Endpoint measurements of the Kurie plot of tritium beta decays, explored at the KATRIN experiment~\cite{Osipowicz:2001sq}, provide an independent probe on the absolute scale of neutrino mass, in terms of the effective mass $m_e$ of the electron neutrino. 
This is complementary to what can be achieved through the cosmological observations considered here. KATRIN currently constrains $m_e < 1.1\,\eV$ at 90\% CL \cite{Aker:2019uuj}, and is expected, in case of a nondetection after 5 years of operation, to establish an upper limit $m_e<0.2\,\eV$ (90\% CL). It is instructive to compare the numbers derived above from current cosmological data, to what would be obtained from KATRIN in the latter case. Assuming the KATRIN nominal 90\% sensitivity, we approximate the posterior $P(\Sigma m_\nu)$ from KATRIN data as a semi-Gaussian peaked in $0$ and with standard deviation $\sigma(\Sigma m_\nu)= 0.36\eV$. In this case, odds of $2:1$ are obtained in favour of nondegenerate masses ($\log_{10} B=-0.3$). 

\begin{figure}
\begin{center}
\includegraphics[width=0.45\textwidth]{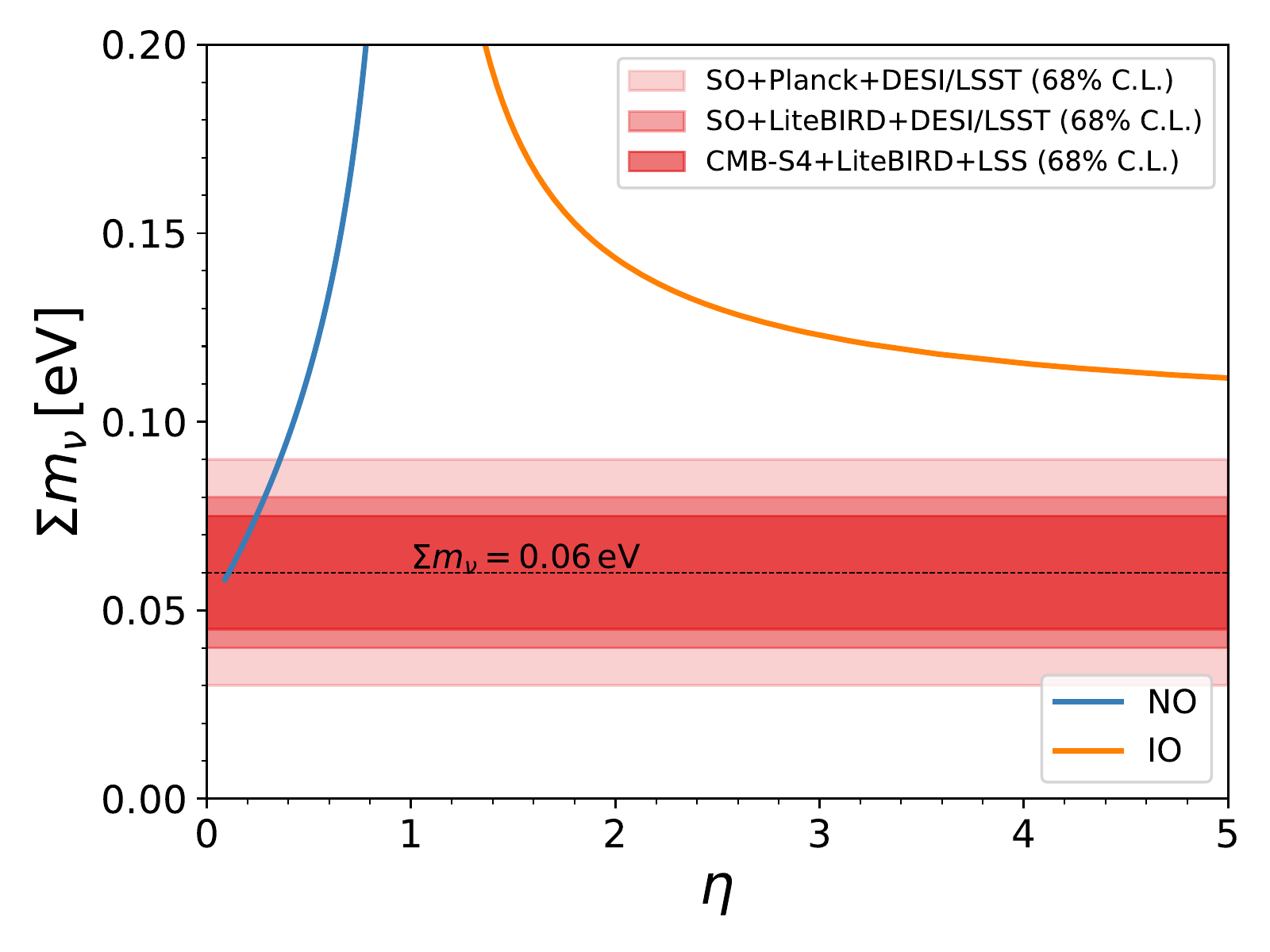}
\includegraphics[width=0.45\textwidth]{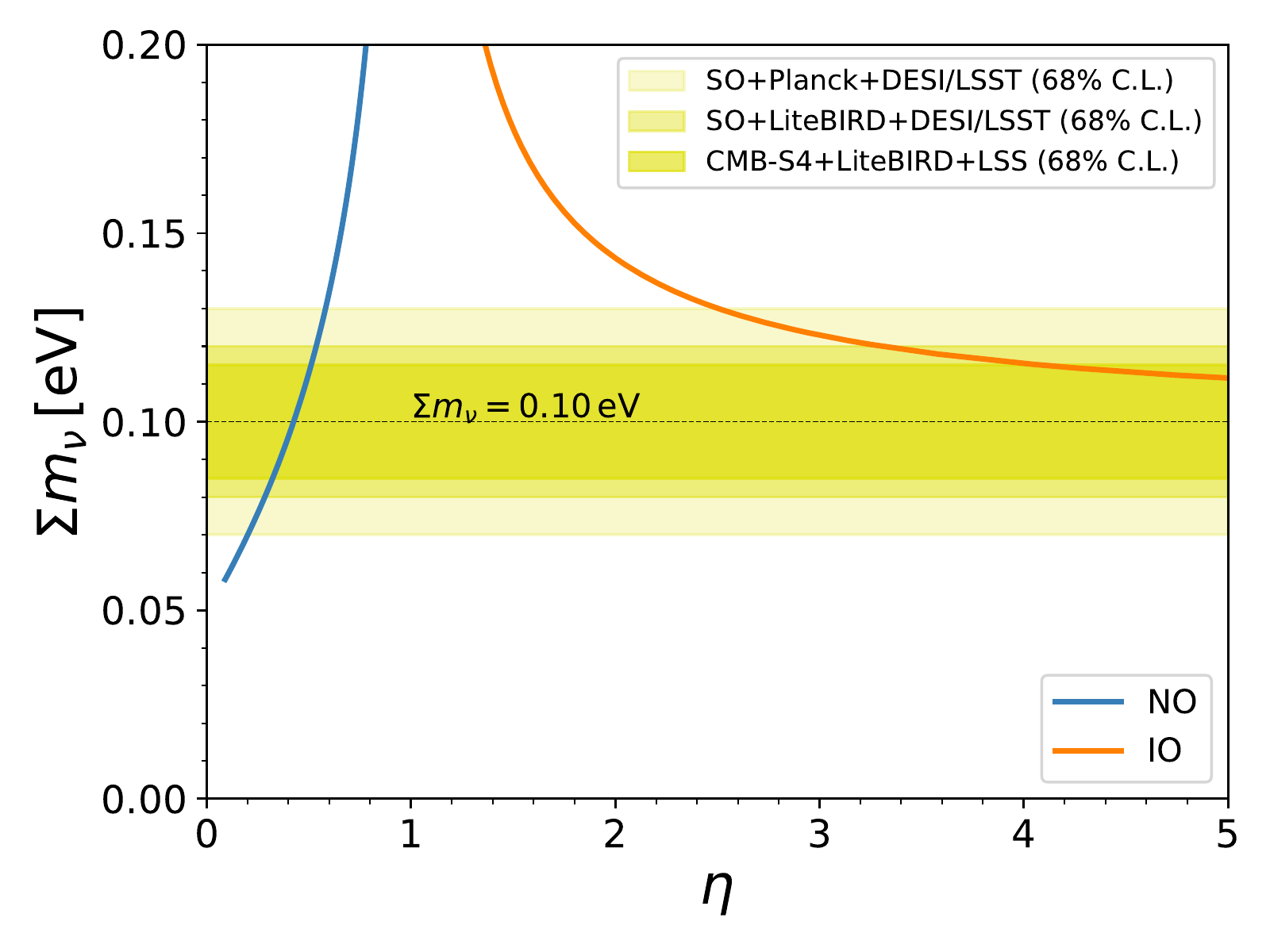}
\caption{Left: Sum of neutrino masses as a function of the parameter $\eta$ (solid curves) in Eqs.~(\ref{eq:eigenvalseta}), given constraints from oscillation experiments, for normal (blue curve) and inverted (orange curve). The three increasingly darker red bands show the 68\% credible interval for $\sum m_\nu$ from future experiments given $\sum m_\nu = 0.06\,\eV$, corresponding to the minimal possible mass for NO, given the results from oscillation experiments. In particular, from the outermost band proceeding to the innermost, we show the expected sensitivity from: Simons Observatory combined with the Planck estimate of $\tau$ and either DESI-BAO or cluster masses calibrated with LSST weak lensing ($\sigma(\sum m_\nu)=0.030\,\eV$), Simons Observatory combined with the LiteBIRD cosmic-variance-limited estimate of $\tau$ and either DESI-BAO or cluster masses calibrated with LSST weak lensing ($\sigma(\sum m_\nu)=0.020\,\eV$), CMB-S4 combined with the LiteBIRD cosmic-variance-limited estimate of $\tau$ and DESI-BAO ($\sigma(\sum m_\nu)=0.015\,\eV$). Right: Same as left panel, but now the three increasingly darker yellow bands show the 68\% credible interval for $\sum m_\nu$ from future experiments given $\sum m_\nu = 0.10\,\eV$, corresponding to the minimal possible mass for IO, given the results from oscillation experiments.\label{fig:Mnufut}}
\end{center}
\end{figure}

Future cosmological data would rule out quasi-degenerate neutrinos at the $17\sigma$ level if $\Sigma m_\nu = 0.06\,\eV$, assuming a sensitivity $\sigma(\Sigma m_\nu) = 0.015 \,\eV$. If, instead, $\Sigma m_\nu = 0.10\,\eV$, a possibility already in mild tension with current data, the same sensitivity would yield an exclusion at the $\sim14 \sigma$ level, still basically ruling out the quasi-degenerate hypothesis. In Fig.~\ref{fig:Mnufut} we compare the sensitivity of future cosmological data to $\Sigma m_\nu$ to the theoretical expectation $\Sigma m_\nu(\eta)$, assuming either $\Sigma m_\nu = 0.06\,\eV$ or $\Sigma m_\nu = 0.10\,\eV$.

\begin{figure}
\begin{center}
\includegraphics[width=0.48\textwidth]{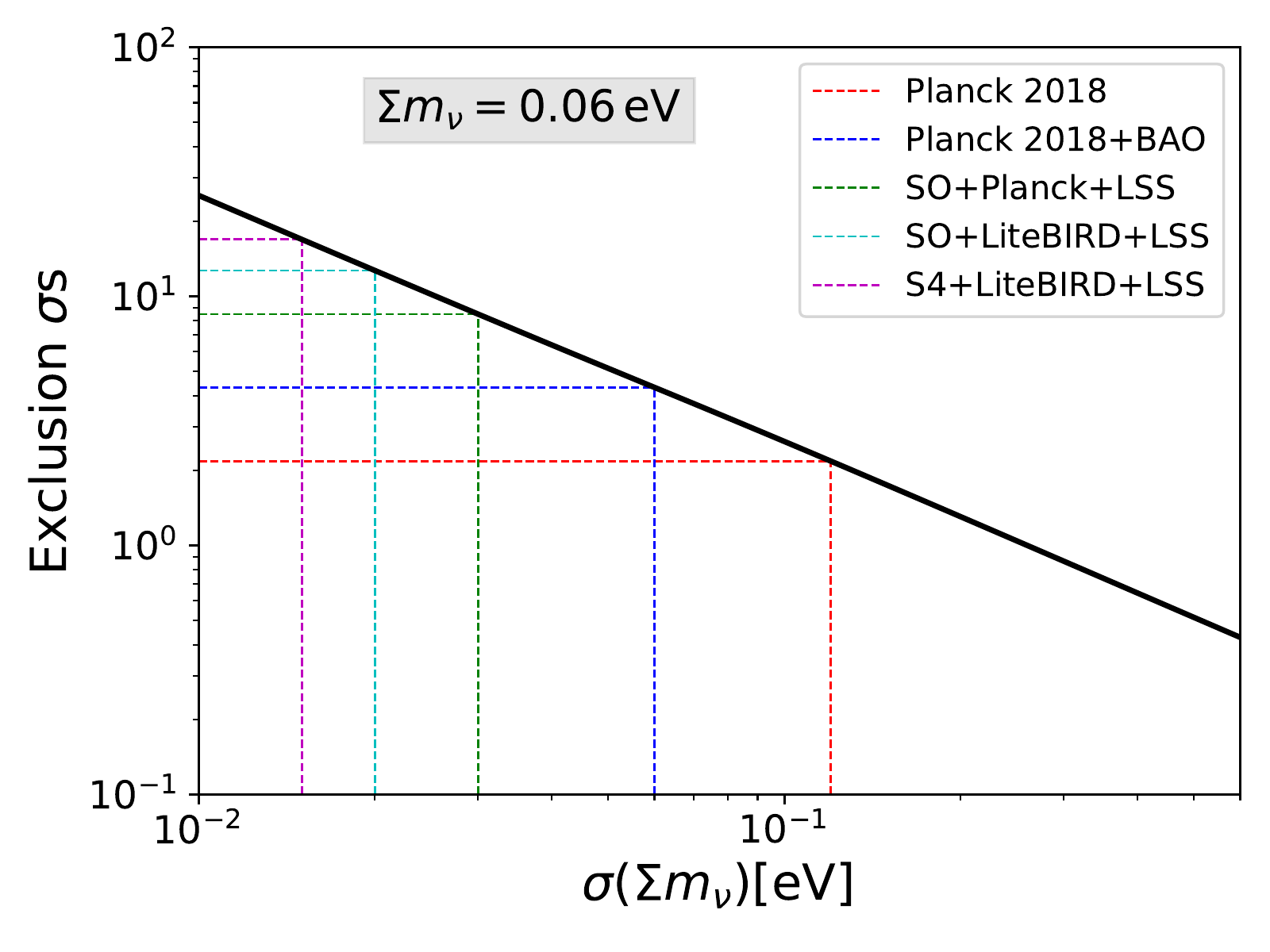}
\caption{
Preference for hierarchical over quasi-degenerate neutrinos, expressed in terms of  Gaussian $\sigma$'s, versus the sensitivity $\sigma(\Sigma m_\nu)$ to the sum of neutrino masses. The dashed lines correspond to the combinations of current and future experiments discussed in the text. The plot assumes the minimal value $\Sigma m_\nu = 60 \,\meV$ allowed by neutrino oscillation experiments for NO. This value yields the highest statistical significance with which one would be able to rule out quasi-degeneracy.
}\label{fig:excl}
\end{center}
\end{figure}

Finally, in Fig.~\ref{fig:excl}, we show the level at which quasi-degenerate neutrinos can be excluded, as a function of the sensitivity $\sigma(\Sigma m_\nu)$, assuming $\Sigma m_\nu = 0.06\,\eV$.

\begin{table}
\begin{center}
\begin{tabular}{|c|c|c|c|}
\hline
 &  Dataset&  $ \eta $ (95\% C.I.) &$ \log_{10} B$\\
\hline
\multirow{2}{*}
{Present} &   Planck 2018&  $ \eta < 0.87 $ or $\eta > 1.2$ & $ -1.8$\\
&   Planck 2018+BAO &  $ \eta < 0.66 $ or $\eta > 1.8$ & $ -8.4 $\\
\hline
\multirow{3}{*}{Future,\,$\Sigma m_\nu=60\,\meV$}& SO+Planck 2018+DESI/LSST ($\sigma(\Sigma m_\nu)=30\,\meV$)&   $ \eta < 0.55 $ or $\eta > 3.0$ & $-16.9$  \\
& SO+LiteBIRD+DESI/LSST ($\sigma(\Sigma m_\nu)=20\,\meV$)&   $ \eta < 0.44 $ or $\eta > 17$ & $-36.9$\\
& CMB-S4+LiteBIRD+DESI ($\sigma(\Sigma m_\nu)=15\,\meV$)&   $\eta<0.35$ & $-64.7$\\
\hline
\multirow{3}{*}{Future,\,$\Sigma m_\nu=100\,\meV$}& SO+Planck 2018+DESI/LSST ($\sigma(\Sigma m_\nu)=30\,\meV$)&   
$ \eta < 0.67 $ or $\eta > 1.8$ & $-12.6$\\
& SO+LiteBIRD+DESI/LSST ($\sigma(\Sigma m_\nu)=20\,\meV$)&   $0.13<\eta<0.61$ or $\eta>2.2$ &$-26.7$\\
& CMB-S4+LiteBIRD+DESI ($\sigma(\Sigma m_\nu)=15\,\meV$)&   $0.20<\eta<0.57$ or $\eta>2.6$ & $-46.3$\\
\hline
\end{tabular}
\end{center}\caption{95\% credible intervals for the parameter $\eta$ from current cosmological data, and projections for future experiments, combined with data from oscillation experiments, in the $(\eta> 0,\,\eta'>0, \,\eta\ge \eta')$ region of parameters. 
Note that oscillation data require that $\eta \ge 0.088$ in this region of the parameter space, so this constraint should be always understood. Constraints in the full parameter space can be reconstructed from symmetry arguments in the $(\eta-\eta')$ plane, see text for details. The constraints in the first row are computed from a combination of the most up-to-date cosmological data available, namely the full Planck 2018 data release, possibly combined with measurements of the baryon acoustic oscillations from BOSS/SDSS. In the following rows we also report forecast results for different combinations of future cosmological surveys, in order of increasing sensitivities: Simons Observatory and CMB-S4 combined with either Planck or LiteBIRD and BAO data from DESI/LSST, see text for details. In case of future surveys, we forecast results assuming that the true value of the summed neutrino masses is either $\sum m_\nu=60\,\meV$ or $\sum m_\nu=100\,\meV$, corresponding to the minimal possible masses for normal and inverted ordering. The constraints are computed according to the prescription discussed in the appendix.
In the last column we also show the logarithm of the Bayes factor between the quasi-degenerate and non-degenerate scenarios $B \equiv P_\mathrm{deg}/(1-P_\mathrm{deg})$. Negative values in that column indicates a preference for non-degenerate neutrinos. The preference gets exponentially larger as the absolute value of $\log_{10} B$ increases.}\label{tab:etafut2}
\end{table}

\subsection{Neutrinoless double beta decay}
\label{sec:neutr-double-decay}

The neutrinoless nuclear double beta decay $(A, Z) \to (A, Z + 2) + 2 e^-$ (denoted $0\nu\beta\beta$) provides another independent and complementary probe of absolute neutrino mass scale, 
especially important as it constitutes a unique model-independent test of the Majorana nature of neutrinos~\cite{Schechter:1981bd}. 

The effective Majorana mass $m_{\beta\beta}$ characterizing $0\nu\beta\beta$ decay is given as
\begin{equation}
m_{\beta\beta}=\left|\sum_jU_{ej}^2m_j\right|=\left|c_{12}^2c_{13}^2m_1+s_{12}^2c_{13}^2m_2e^{2i\phi_{12}}+s_{13}^2m_3e^{2i\phi_{13}}\right|\,,
\end{equation}
where $m_i$ are the neutrino masses, $c_{12}$ and $s_{13}$ correspond to the angles measured from oscillations and $\phi_{12}$, $\phi_{13}$ are the Majorana phases. Note that the amplitude is expressed using the original, symmetrical parametrization of the lepton mixing matrix introduced in Ref.~\cite{Schechter:1980gr}. 
The bounds we have derived above from neutrino oscillation experiments as well as cosmology can be compared also with those that follow from the negative searches for neutrinoless double beta decay~\cite{PhysRevLett.110.062502,KamLAND-Zen:2016pfg,Agostini:2017iyd,Agostini:2018tnm,Agostini:2019hzm,EXO200,PhysRevLett.123.161802,Adams:2018yrj}. 
Current sensitivity should improve significantly in the future, with good prospects for covering the whole region of parameters associated with the inverted ordering spectrum. 
The caveat, though, is that all these $0\nu\beta\beta$ decay bounds rely on nuclear physics calculations of the relevant nuclear matrix elements~\cite{Rodin:2006yk,Simkovic:2019lwa}, which
suffer from non-negligible theoretical uncertainties. For this reason, current bounds on $m_{\beta\beta}$ from $0\nu\beta\beta$ searches are usually expressed as a range of upper limits.

In our study of neutrinoless double beta-decay, we use the full mass matrix of  Sec.~(\ref{subsec:setup2}) with the same $\eta$ as before in Eqns. (\ref{eq:eigenvalseta}).
We compute the effective Majorana mass $m_{\beta\beta}$ as a function of $\eta$, following a similar procedure to the one used for $\Sigma m_\nu$. The individual masses are computed as described in
Sec.~(\ref{subsec:setup2}).
Note that $m_{\beta\beta}$ depends also on the neutrino mixing angles and on the Majorana phases.
As mentioned in the introduction, the scheme we have considered so far, through the mass matrix~(\ref{eq:MnuRe}), implies $\theta_{13}=0$, a value now excluded by oscillation measurements \cite{An:2012eh,Ahn:2012nd}.
However, this can be generalized in order to agree with current oscillation data, as shown in Refs.~\cite{Morisi:2013qna,Chatterjee:2017ilf}, without altering significantly the predictions for neutrino masses. For this reason, we fix the mixing angles to their best-fit values when computing $m_{\beta\beta}$, as we do for the mass splittings.
As far as Majorana phases are concerned, these are treated as free parameters and varied in $\left[0,\,2\pi\right]$.

\begin{figure}
\begin{center}
\includegraphics[width=.7\textwidth]{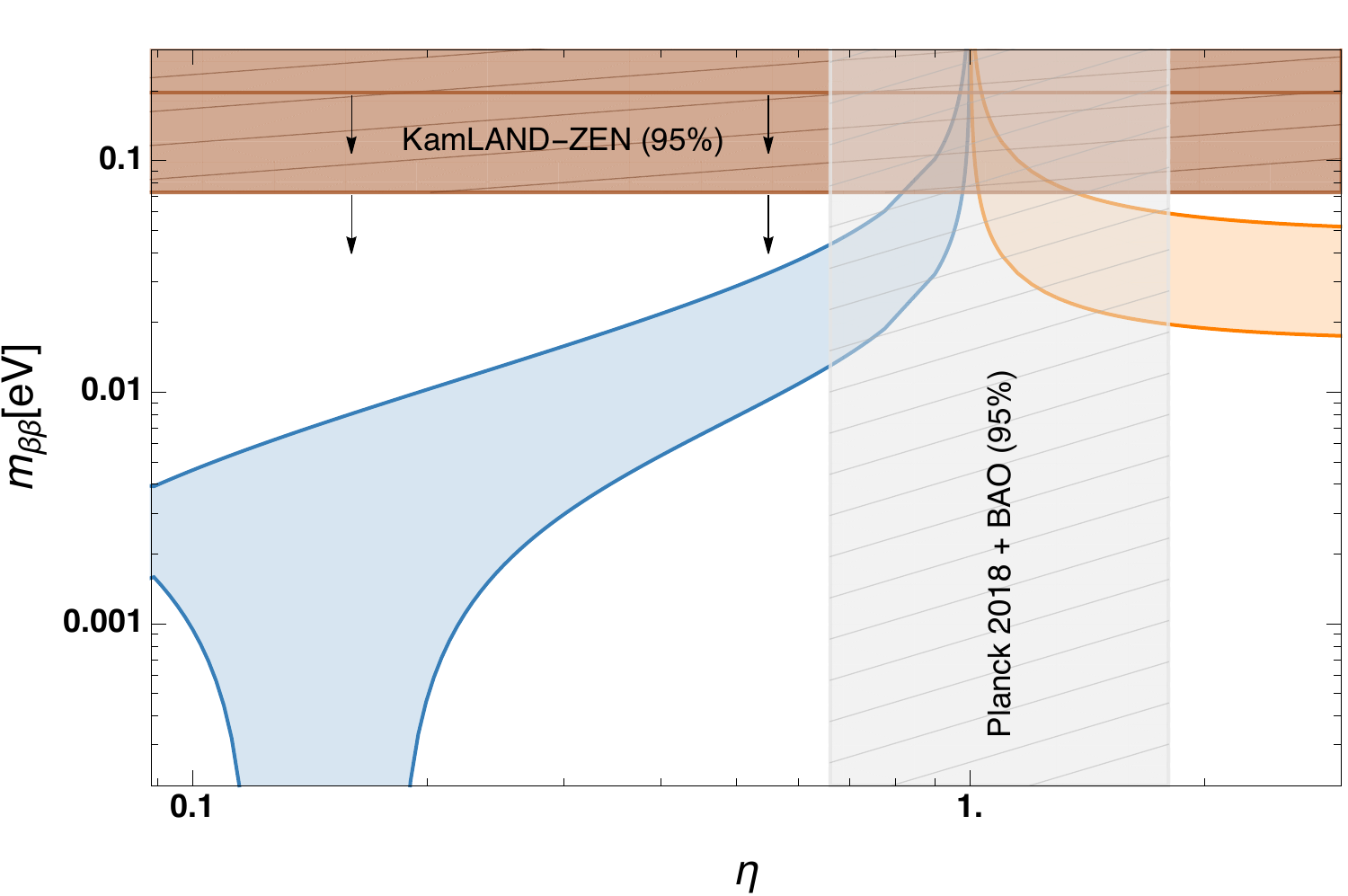}
\caption{Effective Majorana mass $m_{\beta\beta}$ as a function of $\eta$, for normal (blue region) and inverted (orange region) ordering.
The plot assumes $\eta > \eta'$, which implies $\eta \ge 0.088$ to satisfy constraints from oscillation experiments.
  The color code matches the one used in Figs.~(\ref{fig:sumnu_vs_delta_noapprox}) and (\ref{fig:nhihosc}).
  The plot assumes the best-fit values for the mixing angles and mass splittings.
  The width of the bands arises from varying the Majorana phases in $\left[0,\,2\pi\right]$.
  The brown hatched horizontal region corresponds to the 95\% exclusion on $m_{\beta\beta}$ from KamLAND-Zen \cite{KamLAND-Zen:2016pfg}, assuming optimistic nuclear matrix elements.
    The solid line inside the brown region shows the relaxed upper limit obtained for pessimistic nuclear matrix elements.
  The grey hatched vertical band show the 95\% excluded region for $\eta$ from our analysis of Planck 2018 + BAO data in this paper.}
\label{fig:mbb_vs_eta}
\end{center}
\end{figure}

We show in Fig.~(\ref{fig:mbb_vs_eta}) the effective Majorana mass $m_{\beta\beta}$ as a function of $\eta$. Without loss of generality, we restrict our attention to the $\eta>\eta'$ region of the parameter space, as we did when discussing constraints from cosmological data.
Note that, due to the variation of the Majorana phases, a range of theoretical values of $m_{\beta\beta}$
corresponds to a single value of $\eta$.
We also show the upper limits from KamLAND-Zen, that currently provides the most stringent constraints on $m_{\beta\beta}$, i.e., $m_{\beta\beta} < 0.061 - 0.165 \,\eV$ (90\% confidence level)
\cite{KamLAND-Zen:2016pfg}.
We see that the quasi-degenerate region $\eta \simeq 1$ is disfavoured also by current $0\nu\beta\beta$ data, at a level depending on the value assumed for the nuclear matrix elements entering the calculation of the decay amplitude.  For comparison, we also report, in the same figure, the excluded regions for $\eta$ derived in Sec.~(\ref{sec:results}) from current cosmological data.

\section{Conclusions and discussion}
\label{sec:con}

Degenerate neutrino masses can arise via a high-energy symmetry, that is subsequently broken yielding smaller mass splittings. We have considered a mass matrix of the form in Eq.~(\ref{eq:MnuRe}) as a template for this class of models (revamping as in Ref.~\cite{Morisi:2013qna,Chatterjee:2017ilf} is implicit), and derived constraints on the parameters of the matrix using cosmological data together with information from flavour oscillation experiments. A combination of Planck 2018 and BAO data strongly constrains the model parameters, ruling out quasi-degenerate masses at $5.9$ Gaussian $\sigma$'s ($2.4 \sigma$'s if only Planck 2018 CMB data are included). We define the quasi-degenerate scenario as the one in which the large mass splitting is smaller than 10\% of the overall mass scale.

Laboratory experiments also allow us to probe the absolute neutrino mass scale and the elements of the mixing matrix.
We have also compared the constraining power of cosmological data to that of laboratory experiments, in particular KATRIN for $\beta$ decay, and KamLAND-ZEN for $0\nu\beta\beta$ decay.%
The former only provides a weak preference in favour of nondegenerate neutrino masses.
On the other hand, the capability of the latter to provide constraints comparable to those from cosmology is currently hindered by our ignorance of both the matrix elements entering the calculation of the decay amplitude for $0\nu\beta\beta$ as well as the Majorana phases.

The strongest statistical significance with which one could rule out quasi-degeneracy with upcoming experiments is reached if $\Sigma m_\nu = 0.06\,\eV$
(the minimum allowed by oscillation experiments) and assuming a sensitivity $\sigma(\Sigma m_\nu) = 0.015 \,\eV$.
One finds that the exclusion of quasi-degenerate neutrinos from cosmological data would improve to $17$ Gaussian $\sigma$'s.
If, instead, $\Sigma m_\nu = 0.10\,\eV$ (the minimum allowed for inverted ordering), a possibility already in mild tension with current data, the same sensitivity will yield a
$14 \sigma$-level exclusion, still strongly disfavouring the quasi-degenerate hypothesis. If the actual sum of neutrino masses is higher than these minimal values, then 
the quasi-degenerate scenario would still be ruled out, albeit at a slightly lower statistical significance.

  We now discuss how robust are our cosmological bounds,
  commenting briefly on how our conclusions would change by considering a different cosmological model. 
In most extended models, parameter degeneracies act to degrade constraints on neutrino masses.
This is typically the case for models in which the curvature density parameter $\Omega_k$, or the equation of state parameter $w$ of dark energy are allowed to vary.
For example, using a combination of Planck 2015 temperature and low-$\ell$ polarization data and BAO observations, the $95\%$ constraints on $\Sigma m_\nu$ degrade from
$0.19\,\eV$ (\LCDM$+\Sigma m_\nu$) to $0.30\,\eV$ (\LCDM$+\Sigma m_\nu+\Omega_k$) or $0.31\,\eV$ (\LCDM$+\Sigma m_\nu+w$) \cite{Vagnozzi:2017ovm} (see also Ref.~\cite{DiValentino:2019dzu}
for an analysis using Planck 2018 data and using extended models with up to 6 additional parameters).
For such models, the conclusions of this paper -- that basically rely on the cosmological upper limit on $\Sigma m_\nu$ -- would be weakened, allowing a larger portion of the parameter space for
quasi-degenerate neutrinos with respect to \LCDM. In particular, from the upper bounds quoted above one has that quasi-degenerate neutrinos would be disfavoured at the $3.2\sigma$ level in the minimal extension \LCDM+$\Sigma m_\nu$. This rejection would further weaken to the $2.0\sigma$ level if $w$ or $\Omega_k$ are also allowed to vary.
  Note that these numbers should not be directly compared with the results presented in the main text, since they use different data combinations.
  They should, however, illustrate how the level at which quasi-degenerate neutrinos are disfavoured changes when the limits on $\Sigma m_\nu$ are relaxed.
  Figure~\ref{fig:excl} can be used to translate upper bounds on $\Sigma m_\nu$ obtained in an extended model, into an exclusion level for quasi-degenerate neutrinos.
We conclude by stressing that enlarging the parameter space beyond $\Lambda\mathrm{CDM}+\sum m_\nu$ does not always lead to weaker bounds.
As a noticeable example we have models of nonphantom dynamical dark energy (that have $w(z)\geq -1$ at all redshifts $z$), in which the constraints on $\Sigma m_\nu$ are actually
slightly tighter than in \LCDM\ \cite{Vagnozzi:2018jhn}. Thus our conclusions would still hold, and get slightly stronger, in such dark energy models.

\section*{Appendix}

In this appendix we discuss in more detail how the constraints on $\eta$ presented in the main text have been derived.
The starting point is the relation between $(\eta,\,\eta',\, m_0)$ and the sum of neutrino masses $\sum m_\nu$ discussed in Sec.~(\ref{subsec:setup2}).
As explained there we can, without loss of generality, study this relation only in the $(\eta >0,\,\eta' >0)$ region of parameter space, since changing the sign of either $\eta$ or $\eta'$ would
leave $\sum m_\nu$ unchanged. Moreover, exchanging $\eta$ and $\eta'$ also does not change the mass eigenstates, so we can further restrict the analysis to one half of the first quadrant; for definiteness, let us take it to be $\eta \ge \eta'$.
We have used the information from neutrino oscillation experiments to write $\sum m_\nu \Big(\eta,\,\eta',\,m_0\Big)$ as a function of $\eta$ only, i.e.
$\sum m_\nu \Big(\eta,\,\eta'(\eta),m_0(\eta)\Big)\equiv F(\eta)$.
In the first octant of the $(\eta, \eta')$ plane, this relation is one-valued, so that the value of $\eta$ uniquely determines $\Sigma m_\nu$.
The opposite is not true, since all values of $\sumnu$ larger than $0.10\,\eV$ can be obtained from two distinct values of $\eta$ in the first octant.
Note that oscillation data constrain $\eta \ge 0.088$ in the first octant, since a smaller value would yield $\sumnu < 0.06\,\eV$, which is the minimum value allowed by oscillation experiments. The range $\eta\in[0.088,\,\infty)$ can be further split in two regions, corresponding to NO ($\eta <1$) and IO ($\eta >1$). The case $\eta = 1$ corresponds to exact mass degeneracy, which in turn means $\sumnu \to \infty$, while $\eta = 0.088$ and $\eta\to \infty$ yield the minimum masses allowed in NO ($\sumnu = 0.06\,\eV$) and IO ($\sumnu = 0.10\,\eV$), respectively. $F(\eta)$ is always increasing (decreasing) in the NO (IO) region.
Then $F(\eta)$ maps $\eta \in [0.088\,1]$ to $\sumnu \in [0.06,\,\infty)\,\eV$ and $\eta \in [1,\,\infty)$ to $\sumnu \in [0.10,\,\infty)\,\eV$. Note also that the function $F(\eta)$ can be inverted over each of the two sub-ranges separately.
Referring to Fig.~\ref{fig:nhihosc}, we are looking at the half of the blue curve lying in the lower octant, and at the orange curve.

The probability density for $\eta$ is thus obtained from the posterior $P_{\sumnu}\left(\Sigma m_\nu\right)$ of the sum of neutrino masses, provided by cosmological data, as
\begin{equation}
P_\eta(\eta) \propto P_{\sumnu}\Big(\sumnu = F(\eta)\Big) \times \frac{d F(\eta)}{d\eta}\,,
\label{eq:Pofeta}
\end{equation}
up to a proportionality constant that can be obtained a posteriori by requiring that $\int_0^\infty P(\eta) d\eta =1$. Note that the latter requirement amounts to the following normalization for the $\sumnu$ posterior (in the following, it should be understood that $\sumnu$ is measured in $\eV$):
\begin{equation}
\int_{0.06}^{\infty} P_{\sumnu}(\sumnu) d\sumnu + \int_{0.10}^{\infty} P_{\sumnu}(\sumnu) d\sumnu = 1\,.
\end{equation}
This is a consequence of the fact that when $\eta$ varies from $0.088$ to $\infty$, we are traversing the posterior for $\sumnu$ from $0.06$ to $\infty$ and then again from $\infty$ to $0.10$.

The posterior for $\eta$ can be used to compute credible intervals for this parameter. In particular, in the main text we quote 95\% credible intervals. Such an interval $\cal{I}_\eta$ is defined as an interval containing $95\%$ of the total probability:
\begin{equation}
\int_{\cal{I}_\eta} P_\eta(\eta)d\eta  = 0.95\,,
\end{equation}
and is possibly composed by different disconnected regions. The above requirement does not uniquely defines $\cal{I}_\eta$, since there are infinitely many intervals covering 95\% of the total probability. Different prescriptions exist for singling out one of these intervals. We choose to quote for $\eta$ the 95\% interval with the property that the probability for $\sumnu(\eta)$ in every point outside the interval is smaller or equal than the probability inside the interval, i.e.:
\begin{equation}
P_{\sumnu}(F(\eta_1)) \ge  P_{\sumnu}(F(\eta_2)) \quad \textrm{for every} \quad \eta_1 \in {\cal I}_\eta,\quad \eta_2 \notin {\cal I}_\eta
\end{equation}
 This basically amounts to compute the so-called minimum credible interval for $\sumnu$ and map it to the $\eta$ parameter space to get $\cal{I}_\eta$. Note that this is not the same as computing the minimum credible interval for $\eta$, since the reparametrization $\eta \to \sumnu = F(\eta)$ conserves probability mass  but not probability density. In other words, while a 95\% credible interval remains a 95\% credible interval after reparametrization, the condition that probability outside the interval is lower that the probability inside does not necessarily hold in the new parametrization. We choose to compute the minimum credible interval in $\sumnu$ instead than $\eta$ because the former is the parameter that is more directly constrained by cosmological observations and that gets a flat prior in our analysis, so we regard it (observation-wise) as a ``primary'' parameter as opposed to $\eta$, that we regard as a derived parameter.
 
 As noted above, this procedure for constructing the credible interval can result in an interval composed by several disconnected regions, and, in fact, this is nearly always the case in the case under study. The reason is that, in general, there are two regions of large probability, one for each ordering, corresponding to the values of $\eta$ that yield values of the mass close to the minimum values allowed by oscillation experiments for NO and IO. Since these regions are separated in $\eta$ space, corresponding to $\eta \to 0.088$ (NO) and $\eta \to \infty$ (IO), the resulting credible interval is the union of two disconnected regions. This explains the intervals quoted in Tab.~\ref{tab:etafut2}. The only case in which only one integral appears is for future experiments with sensitivity $\sigma(\sumnu) = 0.015 \,\eV$ and fiducial value $\sumnu=0.06\,\eV$. In that case IO is excluded by the data and only the interval corresponding to NO survives.

The fact that the two regions $\eta<1$ and $\eta>1$ correspond to NO and IO also provides a neat way to quantify the preference for one or the other ordering. In particular, one can compute the probabilities for the two orderings, $P_{\mathrm NO}$ and $P_{\mathrm IO}$, as:
\begin{align}
&P_\mathrm{NO} = P_\eta (0.088\le \eta < 1) = \int_{0.088}^1 P_\eta (\eta) d\eta \, ; \\
&P_\mathrm{IO} = 1-P_\mathrm{NO} = P_\eta ( \eta \ge 1) = \int_1^{\infty} P_\eta (\eta) d\eta \, .
\end{align}

We can similarly quantify preference for, or against, degenerate neutrinos from current and future data. In order to do this we need to set a threshold that defines ``quasi-degenerate''. This is somewhat arbitrary; we choose the criterion $|1-\eta|<0.1$. Note that the oscillation constraints ensure that $\eta' \ll \eta$ when this criterion is satisfied. Then, from Eqs~(\ref{eq:eigenvalseta}) it is easy to see that $|1-\eta|$ is the ratio between the large mass splitting and the overall mass scale $m_0$. We then measure the preference for nondegenerate neutrinos by comparing the probabilities enclosed inside and outside the $|1-\eta| < 0.1$ region. In particular we define the probabilty $P_\mathrm{deg}$ for quasi-degenerate neutrinos as:
\begin{equation}
P_\mathrm{deg} \equiv P_\eta (0.9\le \eta \le 1.1)\int_{0.9}^{1.1} P(\eta) d\eta\,,
\end{equation}
and $P_\mathrm{non-deg} = 1 - P_\mathrm{non-deg}$.

Given two mutually excluding scenarios (hypotheses) like NO vs. IO, or quasi-degenerate vs. non-degenerate, the information about the preference for one scenario over the other can be conveyed in different ways. Let us call the two scenarios 1 and 2, with $P_1>P_2$, so that $P_1$ is favoured by the data. A possibility is to directly quote  one of the probabilities $P_1$ and $P_2$. Another possibility is to quote the ratio 
of $P_1$ to $P_2$ in terms of odds, like in ``scenario 1 is favoured with odds 3:2'', meaning that $P_1/P_2 = 3/2$, or $P_1 = 0.6$ and $P_2 = 0.4$. One can similarly quote the so-called Bayes factor $B\equiv P_1/P_2$, so that in the example above $B=1.5$. Finally, in the main text we also translate these probabilities to an equivalent number $x$ of Gaussian standard deviations. This is defined as the value $x$ such that, considering a normal probability distribution with zero mean and unit variance, the probability in the region $-x$ and $+x$ equals $P_1$. This leads to the relation $P_1 = 1-\mathrm{erf}(x/\sqrt{2})$, where $\mathrm{erf}$ is the error function. Note that our use of Gaussian standard deviations should not be meant to imply that we employ frequentist statistics. The statistical analysis presented in this paper is perfomed in the framework of Bayesian statistics.

\begin{acknowledgments}
Work supported by the Spanish grants SEV-2014-0398 and FPA2017-85216-P (AEI/FEDER, UE), PROMETEO/2018/165 (Generalitat Valenciana) and the Spanish Red Consolider MultiDark FPA201790566REDC. MG acknowledges support by Argonne National Laboratory (ANL). Argonne National Laboratory's work was supported under the U.S. Department of Energy contract DE-AC02-06CH11357.
KF is Jeff \& Gail Endowed Chair in Physics at University of Texas, Austin, and is grateful for support. KF acknowledges support by the Vetenskapsr{\aa}det (Swedish Research Council) through contract No. 638-2013-8993 and the Oskar Klein Centre for Cosmoparticle Physics. KF and GK acknowledge support from DoE grant DE-SC007859 and the LCTP at the University of Michigan. ML and MG acknowledge support from INFN through the InDark and Gruppo IV fundings. ML and MG acknowledge support from the ASI grant 2016-24-H.0 COSMOS ``Attivit\`a di studio per la comunit\`a scientifica di cosmologia''. ML thanks the Argonne National Lab for kind hospitality while this work was being finalized. 

\end{acknowledgments}

\bibliographystyle{utphys}
\bibliography{bibliography}
\end{document}